\documentclass[conference,letterpaper]{IEEEtran}

\usepackage[utf8]{inputenc} 
\usepackage[T1]{fontenc}
\usepackage{url}
\usepackage{ifthen}
\usepackage{cite}
\usepackage[cmex10]{amsmath} 

\usepackage{array}
\usepackage{verbatim}
\usepackage{tikz}
\usepackage{epstopdf}
\usepackage{amsfonts}
\usepackage{amsmath}
\usepackage{amssymb}
\newtheorem{defi}{Definition}
\newtheorem{theorem}{Theorem}

\newtheorem{corollary}{Corollary}

\newcommand{\md}{\mathrm{mod}}

\usepackage[center]{caption}
\usepackage{graphicx}
\usepackage{multicol}
\usetikzlibrary{shapes.misc}
\tikzset{cross/.style={cross out, draw=black, minimum size=2*(#1-\pgflinewidth), inner sep=0pt, outer sep=0pt},
cross/.default={1pt}}

\begin{document}
\title{Towards Scalable Security in Interference Channels With Arbitrary Number of Users} 


\author{%
  \IEEEauthorblockN{Parisa Babaheidarian}
  \IEEEauthorblockA{Qualcomm Technologies\\
                    pbabahei@qti.qualcomm.com}
  \and
  \IEEEauthorblockN{Somayeh Salimi}
  \IEEEauthorblockA{Cygate AB \\ somayeh.salimi@cygate.se}
  \and
  \IEEEauthorblockN{Panos Papadimitratos}
  \IEEEauthorblockA{KTH Royal Institute of Technology\\ papadim@kth.se}
}


\maketitle

\begin{abstract}
   In this paper, we present an achievable security scheme for an interference channel with arbitrary number of users. In this model, each receiver should be able to decode its intended message while it cannot decode any meaningful information regarding messages intended for other receivers. Our scheme achieves individual secure rates which scale linearly with $\log(\mathrm{SNR})$ and achieves sum secure rates which is within constant gap of sum secure capacity. To design the encoders at the transmitters side, we combine nested lattice coding, random i.i.d. codes, and cooperative jamming techniques. Asymmetric compute-and-forward framework is used to perform the decoding operation at the receivers. The novelty of our scheme is that it is the first asymptotically optimal achievable scheme for this security scenario which scales to arbitrary number of users and works for any finite-valued SNR. Also, our scheme achieves the upper bound sum secure degrees of freedom of $1$ without using external helpers.
\end{abstract}
\section{Introduction}\label{intro}
Wireless communication channels are susceptible to leakage and interception by illegitimate users. Oftentimes, cryptographic algorithms such as the public key systems (PKI) are used to provide confidentiality. Many of such techniques rely on trapdoor functions whose security are threatened by advances in quantum computers and artificial intelligence. On the other hand, information theoretic tools such as   i.i.d. random codes in \cite{csiszar1978broadcast,wyner1975wire}, promise unconditional security. These techniques have been vastly studied in different communication models including interference channels \cite{liang2009information}. In the last decade, studies showed that despite promising performance of random codes in achieving reliable transmission, these codes perform poorly in achieving high secure rates in high SNR regime. In \cite{xie2014secure, xie2015secure} it was shown that the i.i.d. Gaussian random codes achieve \textit{zero} sum secure degrees of freedom as SNR approaches infinity. To combat this limitation, structured codes have been incorporated in several security scenarios in which they outperformed Gaussian random codes \cite{he2009secure, bagherikaram2010secure,xie2014secure}. 
In \cite{babaheidarian2016security}, Babaheidarian et al., presented an achievable scheme using structured lattice codes which was shown to provide weak secrecy, defined in \cite{maurer2000information}, in a two-user interference channel with weak or moderately weak interference power levels. The advantage of their scheme compared to prior research based on real alignment \cite{xie2014secure} is that the scheme in \cite{babaheidarian2016security} maintains security at any finite SNR value and the secure rates linearly scale with $\log(\mathrm{SNR})$. Furthermore, they showed their scheme is asymptotically optimal. However, the scheme in \cite{babaheidarian2016security} assumed only a two-user scenario and the direct generalization to arbitrary number of users is not straightforward.\\
In this work, we present a new achievable secure scheme for an interference channel with arbitrary number of users specifically ($K>2$) users in which interference level is within weak or moderately weak regimes. Inspired by \cite{ordentlich2014approximate, babaheidarian2015finite, babaheidarian2016security}, our scheme utilizes the compute-and-forward decoding framework to handle finite SNR regimes as opposed to real-alignment schemes in \cite{xie2015secure,bagherikaram2010secure} which applied a maximum likelihood decoder. Our scheme takes advantage of a two-layer codebook structure in which the inner layer uses a set of nested lattice codebooks and the outer layer uses i.i.d. repeated codes. The novelty of our scheme is that the proposed scheme scales to any number of users ($K>2$) and works at any finite SNR value. Also, we show that our scheme achieves optimal sum secure degrees of freedom of $1$ asymptotically. Thus, our achievable sum secure rate is within constant gap from sum secure capacity in finite SNR regime. It is worth to mention that unlike prior schemes in \cite{xie2014secure} and \cite{babaheidarian2017preserving}, in our scheme, transmitters collectively ensure confidentiality of their messages at every unintended receiver without using any external helper.\\
The rest of the paper is organized as follows: Section \ref{probstat} defines the problem statement and our assumptions, Section \ref{mainres} introduces our achievablity results, Section \ref{achieve} provides proof of achievablity and finally conclusion remarks are presented in Section \ref{conclude}.
\section{Problem Statement}\label{probstat}
In this paper, we focus on the problem of simultaneous transmission of secure messages to their intended receivers in a $K$- user interference channel where $K$ is an arbitrary \textit{even} number and $K>2$. For the case with odd number of users, one dummy user is added. At receiver $i$ ($1\leq i \leq K$), the channel output is denoted as $\mathbf{y}_i$ and at transmitter $j$ the input to the channel is denoted as $\mathbf{X}_j$. The channel gain between transmitter $j$ and receiver $i$ is denoted as $h_{ji}$, and lastly, the noise at receiver $i$ is modeled as an i.i.d. random Gaussian vector with zero mean and identity covariance matrix and it is denoted as $\mathbf{z}_i$. The relation between input and output of the channel is defined as
\begin{equation}
\mathbf{y}_i=h_{ii}\mathbf{X}_i+\sum_{j \neq i}h_{ji}\mathbf{X}_j+ \mathbf{z}_i ~~~\forall i \in \{1,\dots, K\}
\end{equation}
\begin{figure}\label{fig1}
\centering
\begin{tikzpicture}[scale=0.65]
\draw(0.7,6) node (nodew1) {$W_1$};
\draw(0.7,3.5) node (nodewi) {$W_i$};
\draw (2,6) node[draw] (nodeE1) {$\mathcal E_1$};
\draw (2,3.5) node[draw] (nodeEi) {$\mathcal E_i$};
\draw (7,6) node[circle,draw,inner sep=0] (nodeplus1) {$+$};
\draw (7,3.5) node[circle,draw,inner sep=0] (nodeplusi) {$+$};
\draw (7,7) node (nodenoise1) {$\mathbf z_1$};
\draw (7,4.5) node (nodenoisei) {$\mathbf z_i$};
\draw (9,6) node[draw] (nodeD1) {$\mathcal D_1$};
\draw (9,3.5) node[draw] (nodeDi) {$\mathcal D_i$};

\draw(10.7,6) node (nodewhat1) {$\begin{array}{c}
 \hat W_1 \end{array} $
 };
 \draw(10.7,3.5) node (nodewhati) {$\begin{array}{c}
 \hat W_i \end{array} $
 };
 \draw(12,6) node (nodewhat4) {$\begin{array}{c}
 \{W_j\}_{j=2}^K \end{array} $
 }; 
 \draw(12,3.5) node (nodewhat4i) {$\begin{array}{c}
 \{W_j\}_{j \neq i} \end{array} $
 }; 
\draw (11.5,6) node[cross=8pt,red] {};
\draw (11.5,3.5) node[cross=8pt,red] {};
\draw[->] (nodew1)--(nodeE1);
\draw[->] (nodewi)--(nodeEi);
\draw[->] (nodeE1)--(nodeplus1) node[pos=0.1,sloped,above]{$\mathbf{X}_1$} node[pos=0.78,sloped,above] {$\scriptstyle \mathrm h_{11}$};
\draw[->] (nodeEi)--(nodeplusi) node[pos=0.1,sloped,above]{$\mathbf{X}_i$} node[pos=0.78,sloped,above] {$\scriptstyle \mathrm h_{ii}$};
\draw[->] (nodenoise1)--(nodeplus1);
\draw[->] (nodenoisei)--(nodeplusi);
\draw[->] (nodeplus1)--(nodeD1) node[pos=0.3,sloped,above]{$\mathbf y_1$};
\draw[->] (nodeplusi)--(nodeDi) node[pos=0.3,sloped,above]{$\mathbf y_i$};
\draw[->] (nodeD1)--(nodewhat1);
\draw[->] (nodeDi)--(nodewhati);
\draw(0.7,1.5) node (nodew3) {$W_K$};
\draw (2,1.5) node[draw, minimum size = 10] (nodeE3) {$\mathcal E_K$};
\draw (7,1.5) node[circle,draw,inner sep=0] (nodeplus3) {$+$};
\draw (7,2.5) node (nodenoise3) {$ \mathbf{z}_2$};
\draw (9,1.5) node[draw] (nodeD3) {$ \mathcal D_K $};
\draw(10.7,1.5) node (nodewhat2) {$\begin{array}{c}
 \hat W_K \end{array} $
 };
\draw(12.25,1.5) node (nodewhat3) {$\begin{array}{c}
 \{W_j\}_{j=1}^{K-1} \end{array} $
 }; 
\draw (11.5,1.5) node[cross=8pt,red] {};
\draw[->] (nodeD3)--(nodewhat2);
\draw[->] (nodew3)--(nodeE3);
\draw[->] (nodeE3)--(nodeplus3)node[pos=0.1,sloped,above]{$\mathbf{X}_K$} node[pos=0.78,sloped,above] {$\scriptstyle h_{KK}$};
\draw[->] (nodenoise3)--(nodeplus3);
\draw[->] (nodeplus3)--(nodeD3) node[pos=0.3,sloped,above]{$\mathbf y_K$};
\draw[->] (3.25,6)--(nodeplus3) node[pos=0.2,below,yshift=-2]{$\scriptstyle \mathrm h_{1K}$};
\draw[->] (3.25,6)--(nodeplusi) node[pos=0.3,below,yshift=13]{$\scriptstyle \mathrm h_{1i}$};
\draw[->] (3.25,3.5)--(nodeplus3) node[pos=0.1,below,yshift=-2]{$\scriptstyle \mathrm h_{iK}$};
\draw[->] (3.25,1.5)--(nodeplus1) node[pos=0.66,above, yshift=3]{$\scriptstyle \mathrm h_{K1}$};
\draw[->] (3.25,1.5)--(nodeplusi) node[pos=0.6,above, yshift=3]{$\scriptstyle \mathrm h_{Ki}$};
\draw[->] (3.25,3.5)--(nodeplus1) node[pos=0.6,above, yshift=3]{$\scriptstyle \mathrm h_{i1}$};
\end{tikzpicture}
\vspace{0.02 in}
\caption{ \small {The $K$-user Gaussian interference channel model with confidential messages.}}
\vspace{-5mm}
\end{figure}
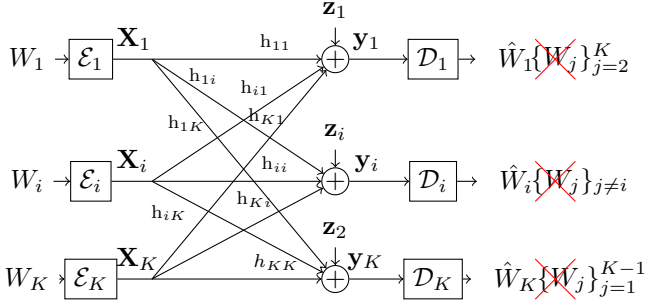
Our assumption is that the channel gains are real valued and known by the transmitters. Fig. 1 illustrates the communication model. We assume that the transmitted and received codewords, i.e., $\mathbf{X}_j$ and $\mathbf{y}_i$ are of length $N$, for all $i,j \in \{1,\dots, K\}$. Transmitter $j$ has an independent confidential message for receiver $j$ which is denoted as $W_j$ and is uniformly distributed over the set $\{1,2, \dots, 2^{NR_j}\}$. Transmitter $j$ encodes its message to codeword $\mathbf{X}_j$ through a stochastic encoder $\mathcal{E}_j$ subject to a power constraint $\|\mathbf{X}_j\|_2^2 \leq NP$, where $P$ is a positive number. Also, receiver $i$ is equipped with decoder $D_i$ which maps codeword $\mathbf{y}_i$ to an estimate of its message: $\hat{W}_i =D_i(\mathbf{y}_i)$.
\begin{defi}[Achievable secure rates]\label{def1}
For the $K$-user ($K>2$)
Gaussian interference channel with independent confidential
messages, a non-negative secure-rate tuple $(R_1, R_2, \dots, R_K)$ is achievable with weak secrecy, if
for any $\epsilon > 0$ and sufficiently large $N$, there exist encoders $\{\mathcal{E}_j\}_{j=1}^{K}$ and decoders $\{D_i\}_{i=1}^{K}$ such that $\forall i,j \{1,2, \dots, K\}$:
\begin{equation}
\mathrm{Prob}\left(D_{i}(\mathbf{y}_{i})\neq W_{i}\right)< \epsilon, \label{eqdef1}
\end{equation}
\begin{equation}\label{eqdef2}
R_{j}\leq \frac{1}{N}H(W_{j}|\mathbf{y}_1, \dots, \mathbf{y}_{j-1},\mathbf{y}_{j+1}, \dots, \mathbf{y}_{K})+\epsilon
\end{equation}
\end{defi}
\section{Main Results}\label{mainres}
In this section, we present the achievable secure rates the defined interference channel model. \\
We define a few notations to present the secure rates in closed form. Assume $R_{comb}^{\ell}$ is an achievable rate at which confidential message $W_{\ell}$ can be reliably decoded at Receiver $\ell$ without any security constraint. Also, assume $P_{\ell,m} \geq 0$ is the power allocated by Transmitter $\ell$ to encode the $m$-th component of confidential message $W_{\ell}$ where total number of components is set to a positive integer $M$. Assume $m^{*}$ is the index of the component with the densest lattice codebook. Additionally, the power allocated by Transmitter $\ell$ to encode the $m$-th component of the jamming codeword is denoted by $P_{\ell,m}^J$. Also, assume $S \subset \{1, 2, \dots, M\}$ where $\frac{|S|}{M} \rightarrow 1$ for large enough $M$. Then, we have
\begin{theorem}[Achievable secure rates]\label{theorm1}
A non-negative rate tuple $(R_1,R_2, \dots, R_K)$ which satisfies the following inequalities is achievable with weak secrecy for the defined interference channel model:
\begin{multline}\label{theorem1eq}
R_{\ell} < R_{comb}^{\ell}- \\ \shoveleft \max_{i \atop i \neq \ell}\left(\log\left(\sum_{m \in S \atop \frac{|S|}{M} \rightarrow 1}^M \frac{h_{\ell, i}^2 P_{\ell,m}+h_{K-\ell+1,i}^2P_{K-\ell+1,m}^J}{h_{K-\ell+1,i}^2P_{K-\ell+1,m^*}^J}\right)\right)
\end{multline}
\end{theorem}
Note that the supremum of all such rates over power allocations $P_{\ell,m}$ and  $P_{\ell,m}^J$, for all $(\ell,m)$, are also achievable so long as the power allocations satisfy the power constraints in Section \ref{achieve} are satisfied.\\
The codebook structure and the encoding-decoding algorithms are described in Section \ref{achieve}. Our achievable scheme utilizes nested lattice codebooks and random i.i.d. repetitions to generate two-layered lattice codewords. Transmitters apply beam-forming operation on message codewords as well as jamming codewords to ensure the security of the confidential messages at any unintended receiver. Note that despite the cooperative jamming scheme, no online communication among the transmitters is required. Proof of reliable decoding and analysis of weak secrecy are presented in the following section.\\
\begin{corollary}\label{cor1}
The optimal sum secure degrees of freedom (s.s.d.f.) of $1$ for an interference channel with arbitrary $K>2$ users is achievable following our scheme for \textit{weak and moderately weak interference power} is $1$, i.e.,
\begin{equation}
s.s.d.f=\frac{\sum_{\ell=1}^K R_{\ell}}{\frac{1}{2}\log(1+P)} \leq 1
\end{equation}
\end{corollary}
Proof of Corollary \ref{cor1} is presented in Subsection \ref{proof}.
\section{Achievability Scheme}\label{achieve}
We prove the achievablity result presented in Theorem 1 by describing the codebook construction followed by the encoding and decoding operations as well as analysis of secrecy.
\subsection{Codebook construction}\label{codebook}
Our codebook and encoding process is based on the idea of passive cooperation among the transmitters. The passive cooperation happens when each transmitter is a sender of its own message but also acts as a helper to protect the confidentiality of another user's message at the illegitimate receivers. For instance, in the $K$-user setting, transmitter 1 helps protecting Transmitter $K$'s message at all the receivers except receiver $K$, and transmitter $2$ does the same job for user $K-1$ and so forth. The reason we call this cooperation passive is that it does not require users to exchange messages among each other so long as they know the channel state and the index of the user they need to help which can be agreed on in the initial acquisition and prior to online secure transmission. \\
Broadly speaking, Transmitter $i$ protect its confidential message with the help of Transmitter $j=K-i+1$ which generates a random jamming codeword which is beam-formed to align with the $i$-th message codeword at every receiver except receiver $i$, simultaneously. Since the same pair of codewords needs to simultaneously get aligned at multiple receivers with different channel gains, perfect alignment between the two codewords is not possible. However, partial alignment across multiple dimensions can occur. It can be shown that if the messages are encoded across large number of dimensions independently, partial alignment can asymptotically approach perfect alignment \cite{motahari2014real}. Therefore, in our codebook construction, we split the confidential messages and jamming signals into large number of independent components. Each component is encoded separately and the superposition of all components are transmitted over the channel.
The codebooks used for encoding confidential messages and the jamming signals form a nested lattice structure in which the jamming codwords are drawn from a finer lattice codebook compared to message codeword. The reason is that the jamming signal needs to get aligned with any possible realization of the confidential message codeword at unintended receivers. A pair of a coarse and a fine lattice sets are used to encode each individual confidential message and jamming signal. Assume that the pair used to encode the $m$-th component of the confidential message $j$ at transmitter $j$ is denoted as $(\Lambda_j^m, \Lambda_{f,j}^m)$. Also, the associated lattice pair used for protecting the $m$-th component of confidential message $K-j+1$ is denoted as $(\Lambda_{J,j}^m, \Lambda_{fJ,j}^m)$. For each component $m$ the following nested lattice relation holds amongst all the lattice sets:\begin{equation*}
\Lambda \subseteq \Lambda_K^m \subseteq \Lambda_{K-1}^m \subseteq \dots \Lambda_1^m \subseteq \Lambda_{J,K}^m \subseteq \dots \Lambda_{J,1}^m \subseteq \Lambda_{f,K}^m
\end{equation*}
\begin{equation}
 \subseteq \dots \Lambda_{f,1}^m \subseteq \Lambda_{Jf,K}^m \subseteq \dots \Lambda_{Jf,1}^m
\end{equation}

The coarse lattice sets are scaled such that their second moments are equal to $\sigma_{m,K}^2, \dots, \sigma_{m,1}^2, \sigma_{Jm,K}^2, \dots, \sigma_{Jm,1}^2$. Also, the fundamental Voronoi region of these coarse lattice sets are denoted as $\mathcal{V}_{i}^m$ for the coarse lattice associated with the $m$-th component of message $i$ and $\mathcal{V}_{J,i}^m$ for the coarse lattice associated with the $m$-th component of the jamming signal generated at transmitter $i$. The center of a corset of the fine lattice $\Lambda_{f,i}^m$ is an $n$-length random vector (lattice word) and is denoted by $\mathbf{t}_{m,i}$. The inner codebook used for encoding the $m$-th component of message $i$ is defined as the union of all realizations of this vector, i.e., $\mathcal{L}_{m,i}\triangleq \{\mathbf{t}_{m,i} | \mathbf{t}_{m,i} \in \mathcal{V}_{i}^m \}$. Similary, the inner codebook for the jamming signal is defined as $\mathcal{L}_{Jm,i}\triangleq \{\mathbf{u}_{m,i} | \mathbf{u}_{m,i} \in \mathcal{V}_{J,i}^m \}$, where $\mathbf{u}_{m,i}$ is also an $n$-length lattice word associated with the center of a corset of fine lattice $\Lambda_{Jf,i}^m$.\\
We use an i.i.d. random repetition of the inner codeword to construct an outer codeword. This step is done to take advantage of Packing Lemma \cite{el2011network} in the proof of secrecy. Consider a probability distribution $P(\mathbf{t}_{m,i})$ over the elements of codebook $\mathcal{L}_{m,i}$. Transmitter $i$ draws $B$ independent realizations of the inner codeword $\mathbf{t}_{m,i}$ according to distribution $P(\mathbf{t}_{m,i})$. These $n$-length lattice words are concatenated to form an $N\triangleq n \times B$ length vector which is a realization of the outer codeword $\bar{\mathbf{t}}_{m,i}$. To construct the corresponding outer codebook, Transmitter $i$ generates $2^{NR_{comb,m}^i}$ realizations of the outer codeword $\bar{\mathbf{t}}_{m,i}$. This outer codebook is denoted as $\mathcal{C}_{m,i}$. Similary, the outer codebook generated to encode the $m$-th component of the jamming signal at Transmitter $i$ is denoted as $\mathcal{C}_{Jm,i}$ and  its outer codeword element is denoted as $\bar{\mathbf{u}}_{m,i}$.\\
The constructed outer codebooks are partitioned to emulate the wiretap code \cite{wyner1975wire}. To do this, Transmitter $i$ randomly partitions codebook $\mathcal{C}_{m,i}$ into $2^{NR_{m,i}}$ bins of equal sizes. Each bin $(m,i)$ is given an index $w_{m,i}$ where $w_{m,i} \in \{1, \dots,  2^{NR_{m,i}} \}$. These indices are essentially independent submessages of the confidential message $W_i$. The transmitter chooses the non-negative rates  $R_{m,i}$ such that $R_i=\sum_{m=1}^{M} R_{m,i}$, where the secure rate $R_i$ is set to
\begin{multline}
R_i\triangleq R_{comb}^i-\\ \shoveleft\max_{\ell \atop \ell \neq i}\left(\log \left(\sum_{m \in S \atop \frac{|S|}{M} \rightarrow 1}^M \frac{h_{i \ell}^2 P_{i,m}+h_{K-i+1,\ell}^2P_{K-i+1,m}^J}{h_{K-i+1,\ell}^2P_{K-i+1,m^*}^J}\right)\right)
\end{multline}
where for all $m, m^* \in M$, and $i \in \{1,2,\dots,K\}$, quantities $P_{i,m}$ and $P_{i,m}^J$ are positive values that represent power allocation among confidential message and jamming signal components, respectively. These quantities are formally defined in Subsection \ref{encode}.\\
Additionally, for each component of the confidential message $i$ and the jamming signal, transmitter $i$ generates a random dither codeword $\bar{\mathbf{d}}_{m,i}$ and $\bar{\mathbf{d}}_{Jm,i}$, respectively. Dithers are drawn uniformly random from the corresponding Voronoi regions, $\mathcal{V}_{m,i}$ and $\mathcal{V}_{Jm,i}$. Dithers are public information and after selection are provided to all parties.
In the following we establish the details of the codebook construction at Transmitter $1$ which protects Transmitter $K$'s confidential message at receivers $1,2, \dots, K-1$. Codebook construction is performed at the other transmitters in a similar fashion.
\subsection{Encoding}\label{encode}
Transmitter $i$ splits the confidential message $w_i \in \{1, \dots 2^{NR_i}\}$ into $M\triangleq T^{2K-2}$ independent sub-messages, where $T$ is a large number. The $m$-th sub-message is denoted as $w_{m,i}\in \{1, \dots,  2^{NR_{m,i}} \}$. To encode this sub-message, Transmitter $i$ randomly picks an outer codeword $\bar{\mathbf{t}}_{m,i}$ from the corresponding outer codebook $\mathcal{C}_{m,i}$. Next, the selected codeword is mixed with a random dither $\bar{\mathbf{d}}_{m,i}$ according to the following equation
\begin{equation}
\tilde{\mathbf{x}}_{m,i}\triangleq [\bar{\mathbf{t}}_{m,i} + \bar{\mathbf{d}}_{m,i}] \mod \Lambda_i^m
\end{equation}
The modular operation in dithering step is done blockwise over each $n$-length block, separately. Similarly, the jamming codeword $\tilde{\mathbf{x}}_{m,i}^J$ is defined. Note that the lattice set associated with the jamming codwords are denser than the message codewords. In the next step, beam-forming operation is performed over each sub-message codeword. Note that each component is sent over a different beam-forming dimension where the number of all the dimensions is $M$. The idea is to align the jamming signal and the confidential message codeword across many such dimensions at unintended receivers. The precoder applied to codeword $\tilde{\mathbf{x}}_{m,i}$ is denoted as $f(m,i,K-i+1, \mathbf{H})$, where $\mathbf{H}=(\mathbf{h}_1, \mathbf{h}_2, \dots, \mathbf{h}_K)$ is the matrix of channel gains from all transmitters to all receivers. The precoder $f$ is a mapping that takes sub-message indices $(m,i)$ and channel gains $\mathbf{H}$ as inputs and outputs a scalar value. This mapping ensures that the resulting codewords are rationally independent for all channel gains expect for a small Lebesgue measure. Transmitter $i$ applies the individual precoders over each component codeword and transmits the superposition codeword $\mathbf{x}_i$ over the channel, where
\begin{equation}
\mathbf{x}_i \triangleq \sum_{m=1}^M \tilde{\mathbf{x}}_{m,i} f(m,i,K-i+1,\mathbf{H})
\end{equation} 
Similarly, a precoder is applied to the jamming codeword to protect the confidential message of user $K-i+1$, i.e.,
\begin{equation}
\mathbf{x}_i^J \triangleq \sum_{m=1}^M \tilde{\mathbf{x}}_{m,i}^J g(m,i,K-i+1,\mathbf{H})
\end{equation} 
The transmitted codeword is denoted as $\mathbf{X}_i\triangleq \mathbf{x}_i + \mathbf{x}_i^J$ and it satisfies the power constraint. Let us define $P_{m,i}\triangleq \sigma_{m,i}^2 |f(m,i,K-i+1,\mathbf{H})|^2$ and $P_i=\sum_{m=1}^M P_{m,i}$. Similarly, for the jamming codeword, define $P_{m,i}^J\triangleq \sigma_{Jm,i}^2 |f(m,i,K-i+1,\mathbf{H})|^2$ and $P_i^J=\sum_{m=1}^M P_{m,i}^J$. Transmitter $i$ allocates power between jamming power and message power such that $P_i+P_i^J \leq P$. Furthermore, the coarse lattice sets associated with every jamming codeword is scaled such that for $ \forall i \in \{1,\dots, K\}$, we have
\begin{equation}\label{powerconstJam}
h_{K-i+1,i}^2 P_{K-i+1}^J \leq 1
\end{equation}
Note that the above condition is essential to achieve sum secure degrees of freedom of $1$ and without this condition, the achievable sum secure degrees of freedom would reach $\frac{K}{K+1}$.
The precoder mapping $f(m,i,K-i+1, \mathbf{H})$  is a product of powers of channel gains between both Transmitter $i$ and Transmitter $K-i+1$ and the receivers, i.e.,
\begin{equation*}\small{
f(m,i,K-i+1,\mathbf{H})=(h_{i1}^{r_1}h_{i2}^{r_2}\dots h_{i,i-1}^{r_{i-1}}h_{i,i+1}^{r_{i}}\dots h_{iK}^{r_{K-1}})\times}
\end{equation*}
\begin{equation}
\small{
(h_{K-i+1,1}^{r_K}h_{K-i+1,2}^{r_{K+1}}\dots h_{K-i+1,i-1}^{r_{K+i-1}}h_{K-i+1,i+1}^{r_{K+i}}\dots h_{K-i+1,K}^{r_{2K-2}})}
\end{equation}
and 
\begin{equation*}
\tiny{
g(m,i,K-i+1,\mathbf{H})=(h_{i1}^{r_1}\dots h_{i,K-i}^{r_{K-i}}h_{i,K-i+2}^{r_{K-i+1}}\dots h_{iK}^{r_{K-1}})\times}
\end{equation*}
\begin{equation}
\small{
(h_{K-i+1,1}^{r_K}h_{K-i+1,2}^{r_{K+1}}\dots h_{K-i+1,K-i}^{r_{K-i}}h_{K-i+1,K-i+2}^{r_{K-i+1}}\dots h_{K-i+1,K}^{r_{2K-2}})}
\end{equation}
The powers $(r_1,r_2, \dots, r_{2K-2})$ are computed using a one-to-one mapping $\phi(m)$ that takes the $m$-th beam-forming dimension to the $2K-2$-length tuple power where each power is one of the possible $T$ dimensions. In other words, we have:
\begin{equation}
\phi(m):\{1,\dots,M\} \rightarrow \{1,\dots,T\}\times \{1, \dots, T\} \times \dots \{1,\dots, T\}
\end{equation}
and for every $m \in \{1,\dots, M\}$ there exists a non-negative $2K-2$ length tuple such that
\begin{equation}
(r_1,r_2, \dots, r_{2K-2})=\phi(m) ~r_j \in \{1,2, \dots, T\}
\end{equation}
\subsection{Decoding}\label{decod}
Decoding at each receiver follows asymmetric compute-and-forward technique used in \cite{babaheidarian2017preserving}. In the following, we describe the decoding process at Receiver $i$. Other receivers act in a similar manner.\\
Receiver $i$ observes the scaled lattice codeword associated with its own message plus a set of unintended codewords aligned with jamming codewords plus effective noise as
\begin{multline} \label{yi}
\mathbf{y}_i=h_{ii}\mathbf{x}_i+\sum_{\ell=1 \atop \ell \neq i}^K (h_{\ell i}\mathbf{x}_{\ell} +h_{K-\ell+1,i}\mathbf{x}_{K-\ell+1}^J) \\ +h_{K-i+1,i}\mathbf{x}_{K-i+1}^J +\mathbf{z}_i
\end{multline}
Due to asymptotic alignment \cite{motahari2014real} along many beam-forming dimensions, the collections of the confidential and the jamming codewords participating in the second term in (\ref{yi}) are mutually aligned. Also, note that due to the constraint in (\ref{powerconstJam}), the power of the third term falls below noise power (assuming all noise powers are normalized). Also, this term includes only a jamming signal which is of no use to Receiver $i$. Note that the condition in (\ref{powerconstJam}) is aligned with weakly and moderately weak interference definition in \cite{babaheidarian2016security}. Therefore, Receiver $i$ treats the third term as an additional noise term and the normalized effective noise term $\tilde{\mathbf{z_i}}$ is defined as
\begin{equation}
\tilde{\mathbf{z_i}} \triangleq \frac{1}{\sqrt{1+h_{K-i+1,i}^2P_{K-i+1}^J}} (h_{K-i+1,i}\mathbf{x}_{K-i+1}^J+\mathbf{z_i})
\end{equation}
As a result, Receiver $i$ effectively observes a $K$-user Multiple Access Channel (MAC) at its end, i.e.,
\begin{multline}\label{macEff}
\tilde{\mathbf{y}}_i\triangleq \frac{h_{ii}}{\sqrt{1+h_{K-i+1,i}^2P_{K-i+1}^J}}\mathbf{x}_i \\+ \frac{1}{\sqrt{1+h_{K-i+1,i}^2P_{K-i+1}^J}}\sum_{\ell=1 \atop \ell \neq i}^K (h_{\ell i}\mathbf{x}_{\ell} +h_{K-\ell+1,i}\mathbf{x}_{K-\ell+1}^J) + \tilde{\mathbf{z_i}}
\end{multline}
The effective MAC channel gain vector at Receiver $i$ is denoted as $\mathbf{h}_{eff,i}$ and it is defined as $ \tiny{
\mathbf{h}_{eff,i}\triangleq \left  (\frac{h_{ii}}{\sqrt{1+h_{K-i+1,i}^2P_{K-i+1}^J}}, \frac{1}{\sqrt{1+h_{K-i+1,i}^2P_{K-i+1}^J}},\dots, \right .}$ $ \tiny{ \left .\frac{1}{\sqrt{1+h_{K-i+1,i}^2P_{K-i+1}^J}}\right )^T } $. \\
The ratio between the power of each effective codeword in the effective MAC equation (\ref{macEff}) and the power constraint $P$ is defined as power scaling vector $\mathbf{b}_{eff,i}$ and $ \tiny{
 \mathbf{b}_{eff,i}\triangleq \left (\sqrt{ \frac{P_i}{P}}, \sqrt{\frac{h_{1i}^2P_1+h_{Ki}^2P_K^J}{P}}, \dots, \sqrt{\frac{h_{Ki}^2P_K+h_{1i}^2P_1^J}{P}} \right)^T}
$
 Now, Receiver $i$ applies the compute-and-forward technique used for a MAC channel in \cite{ordentlich2014approximate}. Receiver $i$ finds the nearly optimal set of linear independent integer-valued coefficient vectors which maximize the achievable MAC sum-rate for that Receiver. The receiver constructs $K$-linearly independent equations using these integer-valued coefficient vectors and decode each equation successively. The first equation is to aim to decode the effective lattice codeword with the highest data rate, i.e., the lattice codeword that belongs to the densest fine lattice set. Upon decoding the codeword, it is canceled out from the second equation and so forth. Therefore, the least achievable equation rate is associated with the $K$-th combination equation.\\
Let us denote the optimal set of integer-valued coefficient vectors that construct the $K$ combination equations with $\mathbf{a}_1, \mathbf{a}_2, \dots, \mathbf{a}_K$. Also, let us denote the rates at which the combination equations are decoded at Receiver $i$ as $R_{comb,1}^i, R_{comb,2}^i, \dots, R_{comb,K}^i$. The set of coefficient vectors is computed such that the rate at which combination equations are decoded is non-increasing, i.e., $R_{comb,1}^i\geq R_{comb,2}^i \geq \dots\geq R_{comb,K}^i$ . Basically, the effective codeword with the highest achievable rate is decoded first and canceled out and then the second highest rate effective codeword is decoded and so forth. Let us denote the first combination equation as $\mathbf{v}_1 \triangleq \mathbf{a}_1(\ell) \mathbf{x}_{eff, \ell} $. For instance, the effective codewords at Receiver $i$ are defined as $\mathbf{x}_{eff, 1}\triangleq \mathbf{x}_i$, $\mathbf{x}_{eff, 2}\triangleq h_{1i}\mathbf{x}_{1} +h_{Ki}\mathbf{x}_{K}^J$ and so forth so it would match the corresponding gain order in the effective MAC observed by Receive $i$. Decoding equation $\mathbf{v}_1$ is performed using the compute-and-forward technique \cite{ordentlich2014approximate, nazer2011compute} by scaling the noisy observation and canceling out the public dithers as
\begin{equation}
\beta_1 \tilde{\mathbf{y}}_i- \sum_{\ell=1}^K \mathbf{a}_1(\ell) \bar{\mathbf{d}}_{eff,\ell}= \mathbf{v}_1+ \mathbf{z}_{eff, 1}
\end{equation}
where 
\begin{equation}\label{zeff1}
\mathbf{z}_{eff, 1} \triangleq  \sum_{\ell=1}^K (\beta_1 \mathbf{h}_{eff}(\ell)-\mathbf{a}_1(\ell)) \mathbf{x}_{eff,\ell} +\beta_1 \tilde{\mathbf{z}}_i
\end{equation}
Let us denote the second moment of the effective noise term associated with combination equation $1$ in (\ref{zeff1}) as $\sigma_{eff,1}^2$. Following Theorem $2$ in \cite{nazer2011compute} equation $\mathbf{v}_1$ can be decoded at an achievable rate $R_{comb,1}^i=\frac{1}{2}\log(\frac{P_{eff,j}}{\sigma_{eff,1}^2})$, where $j$ is the index of the effective lattice codeword with densest lattice sets among participating codewords in combination equation $\mathbf{v}_1$. Similarly, combination equations $\mathbf{v}_2, \dots, \mathbf{v}_K$ are constructed and decoded. Assume that the mapping between effective codeword indices and the order at which they get decoded at Receiver $i$ is determined by a one-to-one permutation function $\pi^i(.): \{1, 2, \dots, K\} \rightarrow \{1, 2, \dots, K\}$. Therefore, the achievable combination rates are derived as
\begin{equation}
R_{comb,\ell}^i=\frac{1}{2}\log \left(\frac{P_{eff,\pi^i(\ell)}}{\sigma_{eff,\ell}^2}\right)
\end{equation}
We already established that the combination equations are constructed such that the codeword with the densest lattice set gets decoded first therefore $\sigma_{eff,1}^2 \leq \sigma_{eff,2}^2 \leq \dots \sigma_{eff,K}^2$. As a result, the lowest achievable rate to decode effective codeword $\mathbf{x}_{eff,1}$ at Receiver $i$, i.e., $\mathbf{x}_i$ is
\begin{equation}
R_{comb}^i \triangleq \frac{1}{2}\log \left(\frac{P_{eff,1}}{\sigma_{eff,i,K}^2}\right)=\frac{1}{2}\log \left(\frac{P_i}{\sigma_{eff,i,K}^2}\right)
\end{equation}
Similarly, the achievable combination rates are determined at the other receivers. Since $R_i \leq R_{comb}^i$ for $i \in \{1, \dots, K\}$, the achievability proof is completed. In the following subsection, we show that $R_i$ is achievable with \textit{weak secrecy}. 
\subsection{Analysis of Security}\label{aos}
So far we have shown that the lower bound stated in Theorem \ref{theorm1} are achievable in terms of reliably getting decoded at intended receivers. In this part, we proceed with showing the presented rates provide weak secrecy for every confidential message at all unintended receivers. To do this, it suffice to show the following for an arbitrary Receiver $i \in \{1, \dots, K\}$:
\begin{equation}
\frac{1}{N}I(W_1, \dots, W_{i-1}, W_{i+1}, \dots, W_K; \mathbf{y}_i) \leq \epsilon
\end{equation}
in which $\epsilon > 0$  approaches to zero as $N=nB$ tends to infinity. Note that $\frac{1}{N}I(W_1, \dots, W_{i-1}, W_{i+1}, \dots, W_K; \mathbf{y}_i)= \sum_{\ell=1 \atop \ell \neq i}^K R_{\ell} - \frac{1}{nB}H(W_1, \dots, W_{i-1}, W_{i+1}, \dots, W_K | \mathbf{y}_i)$ and since conditioning does not increase entropy, we have
\begin{multline}\label{aoseqmain}
\frac{1}{N}I(W_1, \dots, W_{i-1}, W_{i+1}, \dots, W_K; \mathbf{y}_i) \leq \\ \sum_{\ell=1 \atop \ell \neq i}^K R_{\ell} - \frac{1}{nB}H(W_1, \dots, W_{i-1}, W_{i+1}, \dots, W_K | \mathbf{y}_i, \bar{\mathbf{t}}_i)
\end{multline}
Next step is to obtain a lower bound on the second term in (\ref{aoseqmain}). Let us define notation $\{\bar{\mathbf{t}}\}_{\ell=1 \atop \ell \neq  i}^K$ which represents the set of outer lattice codewords of all transmitters expect Transmitter $i$. We have
\begin{multline}
 \frac{1}{nB}H(W_1, \dots, W_{i-1}, W_{i+1}, \dots, W_K | \mathbf{y}_i, \bar{\mathbf{t}}_i)\\ =\frac{1}{nB}H(W_1, \dots, W_{i-1}, W_{i+1}, \dots, W_K, \{\bar{\mathbf{t}}\}_{\ell=1 \atop \ell \neq  i}^K, | \mathbf{y}_i, \bar{\mathbf{t}}_i)  \\ -\frac{1}{nB}H(\{\bar{\mathbf{t}}\}_{\ell=1 \atop \ell \neq  i}^K | \mathbf{y}_i, \bar{\mathbf{t}}_i, W_1, \dots, W_{i-1}, W_{i+1}, \dots, W_K) \\ \shoveleft \geq \frac{1}{nB}H( \{\bar{\mathbf{t}}\}_{\ell=1 \atop \ell \neq  i}^K | \mathbf{y}_i, \bar{\mathbf{t}}_i) \\ -\frac{1}{nB}H(\{\bar{\mathbf{t}}\}_{\ell=1 \atop \ell \neq  i}^K | \mathbf{y}_i, \bar{\mathbf{t}}_i, W_1, \dots, W_{i-1}, W_{i+1}, \dots, W_K) \\ \shoveleft{\stackrel{(a)}{\geq} \frac{1}{nB}H(\{\bar{\mathbf{t}}\}_{\ell=1 \atop \ell \neq  i}^K | \mathbf{y}_i, \bar{\mathbf{t}}_i) -2\epsilon_i} \\ 
 \shoveleft{\stackrel{(b)}{\geq} \frac{1}{nB}H(\{\bar{\mathbf{t}}\}_{\ell=1 \atop \ell \neq  i}^K | \mathbf{y}_i, \bar{\mathbf{t}}_i, \{\bar{\mathbf{d}}\}_{\ell=1}^K, \mathbf{x}_{K-i+1}^J,  \mathbf{z}_i) -2\epsilon_i} \\ \shoveleft{= \frac{1}{nB} H(\{\bar{\mathbf{t}}\}_{\ell=1 \atop \ell \neq  i}^K | \sum_{\ell=1 \atop \ell \neq i}^K (h_{\ell i}\mathbf{x}_{\ell} +h_{K-\ell+1,i}\mathbf{x}_{K-\ell+1}^J),\bar{\mathbf{t}}_i,}  \\ \{\bar{\mathbf{d}}\}_{\ell=1}^K, \mathbf{x}_{K-i+1}^J,  \mathbf{z}_i) -2 \epsilon_i \\ 
\shoveleft{ =\frac{1}{nB} H(\{\bar{\mathbf{t}}\}_{\ell=1 \atop \ell \neq  i}^K | \sum_{\ell=1 \atop \ell \neq i}^K (h_{\ell i}\mathbf{x}_{\ell} +h_{K-\ell+1,i}\mathbf{x}_{K-\ell+1}^J),\bar{\mathbf{t}}_i,}  \\ \{\bar{\mathbf{d}}\}_{\ell=1}^K, \mathbf{x}_{K-i+1}^J,  \mathbf{z}_i) -2 \epsilon_i \\
 \tiny{=\frac{1}{nB} H(\{\bar{\mathbf{t}}\}_{\ell=1 \atop \ell \neq  i}^K |\sum_{\ell=1 \atop \ell \neq i}^K \bigg[\sum_{m=1}^M h_{\ell i}\mathbf{x}_{\ell,m} +h_{K-\ell+1,i}\mathbf{x}_{K-\ell+1,m}^J)\bigg] }\\\md ~\Lambda_{m^*,\ell}^J, \sum_{\ell=1 \atop \ell \neq i}^KQ_{\Lambda_{m^*,\ell}^J}\bigg(\sum_{m=1}^M h_{\ell i}\mathbf{x}_{\ell,m} +h_{K-\ell+1,i}\mathbf{x}_{K-\ell+1,m}^J)\bigg),\\\bar{\mathbf{t}}_i, \{\bar{\mathbf{d}}\}_{\ell=1}^K, \mathbf{x}_{K-i+1}^J,  \mathbf{z}_i) -2 \epsilon_i \\
\hspace{-0.1in}\shoveleft{\stackrel{(c)}{\geq}\frac{1}{nB} H\bigg(\{\bar{\mathbf{t}}\}_{\ell=1 \atop \ell \neq  i}^K \bigg |} \\ \sum_{\ell=1 \atop \ell \neq i}^K \bigg[\sum_{m \in S \atop \frac{|S|}{M} \rightarrow 1}^M \left(h_{\ell i}\mathbf{x}_{\ell,m} +h_{K-\ell+1,i}\mathbf{x}_{K-\ell+1,m}^J)\right)\bigg]~\md~\Lambda_{m^*,\ell}^J, \\ \bar{\mathbf{t}}_i,  \{\bar{\mathbf{d}}\}_{\ell=1}^K, \mathbf{x}_{K-i+1}^J,  \mathbf{z}_i\bigg) - \\ \frac{1}{nB}  \sum_{\ell=1 \atop \ell \neq i}^K H \bigg( Q_{\Lambda_{m^{*}_{\ell}}^J}\bigg( \sum_{m \in S \atop \frac{|S|}{M} \rightarrow 1}^M \left(h_{\ell i}\mathbf{x}_{\ell,m} +h_{K-\ell+1,i}\mathbf{x}_{K-\ell+1,m}^J\right)\bigg)\\ \shoveleft - 2 \epsilon_i \\
 \stackrel{(d)}{=} \frac{1}{nB} H\bigg(\{\bar{\mathbf{t}}\}_{\ell=1 \atop \ell \neq  i}^K |\sum_{\ell=1 \atop \ell \neq i}^K\bigg[\sum_{m \in S \atop \frac{|S|}{M} \rightarrow 1}^M h_{\ell i}f(m,\ell,K-\ell+1,\mathbf{H})\bar{\mathbf{t}}_{\ell,m} \\ +h_{K-\ell+1,i}g(m,K-\ell+1,K-\ell, \mathbf{H})\bar{\mathbf{u}}_{K-\ell+1,m}^J\bigg]~\md~\Lambda_{m^*,\ell}^J, \\ \bar{\mathbf{t}}_i, \{\bar{\mathbf{d}}\}_{\ell=1}^K, \mathbf{x}_{K-i+1}^J,  \mathbf{z}_i\bigg) -\\ \frac{1}{nB}  \sum_{\ell=1 \atop \ell \neq i}^K H \bigg( Q_{\Lambda_{m^{*}_{\ell}}^J}\bigg( \sum_{m \in S \atop \frac{|S|}{M} \rightarrow 1}^M \left(h_{\ell i}\mathbf{x}_{\ell,m} +h_{K-\ell+1,i}\mathbf{x}_{K-\ell+1,m}^J\right)\bigg) \\ \shoveleft-2 \epsilon_i  \\
 \stackrel{e}{=} \frac{1}{nB} H\bigg(\{\bar{\mathbf{t}}\}_{\ell=1 \atop \ell \neq  i}^K |\sum_{\ell=1 \atop \ell \neq i}^K\bigg[\sum_{m \in S \atop \frac{|S|}{M} \rightarrow 1}^M h_{\ell i}f(m,\ell,K-\ell+1,\mathbf{H})\bar{\mathbf{t}}_{\ell,m} \\ +h_{K-\ell+1,i}g(m,K-\ell+1,K-\ell, \mathbf{H})\bar{\mathbf{u}}_{K-\ell+1,m}^J\bigg] \mod \Lambda_{m^*,\ell}^J \bigg) \\- \frac{1}{nB}  \sum_{\ell=1 \atop \ell \neq i}^K H \bigg( Q_{\Lambda_{m^{*}_{\ell}}^J}\bigg( \sum_{m \in S \atop \frac{|S|}{M} \rightarrow 1}^M \left(h_{\ell i}\mathbf{x}_{\ell,m} +h_{K-\ell+1,i}\mathbf{x}_{K-\ell+1,m}^J\right)\bigg) \\ \shoveleft-2 \epsilon_i \\ 
\shoveleft{\stackrel{f}{=} \frac{1}{nB} H(\{\bar{\mathbf{t}}\}_{\ell=1 \atop \ell \neq  i}^K) -}\\ \frac{1}{nB}  \sum_{\ell=1 \atop \ell \neq i}^K H \bigg( Q_{\Lambda_{m^{*}_{\ell}}^J}\bigg( \sum_{m \in S \atop \frac{|S|}{M} \rightarrow 1}^M \left(h_{\ell i}\mathbf{x}_{\ell,m} +h_{K-\ell+1,i}\mathbf{x}_{K-\ell+1,m}^J\right)\bigg)\\ \shoveleft- 2 \epsilon_i \\
  \shoveleft{\stackrel{g}{=} \frac{1}{nB} H(\{\bar{\mathbf{t}}\}_{\ell=1 \atop \ell \neq  i}^K)} \\ \shoveleft- \sum_{\ell=1 \atop \ell \neq i}^K \log\left(\sum_{m \in S \atop \frac{|S|}{M} \rightarrow 1}^M \frac{h_{\ell i}^2 P_{\ell,m}+h_{K-\ell+1,i}^2P_{K-\ell+1,m}^J}{h_{K-\ell+1,i}^2P_{K-\ell+1,m^*}^J}\right)\\ \shoveleft- 2\epsilon_i - \delta \\
 \shoveleft{ \stackrel{h}{=} \sum_{\ell=1 \atop \ell \neq  i}^K R_{comb}^{\ell} -} \\ \shoveleft \sum_{\ell=1 \atop \ell \neq i}^K \log\left(\sum_{m \in S \atop \frac{|S|}{M} \rightarrow 1}^M \frac{h_{\ell i}^2 P_{\ell,m}+h_{K-\ell+1,i}^2P_{K-\ell+1,m}^J}{h_{K-\ell+1,i}^2P_{K-\ell+1,m^*}^J}\right) \\\shoveleft{- 2\epsilon_i - \delta}\\
\end{multline}
In the above inequalities the followings hold: inequality (a) is due to i.i.d. random repetitions of the outer codewords and is deduced by using Packing Lemma as shown in \cite{el2011network}. Inequality (b) holds since conditioning does not increase entropy. Also, we defined the set of all public dither codewords from all users with notation $\{\bar{\mathbf{d}}\}_{\ell=1}^K$. Inequality (c) is concluded from the principal that joint entropy is not bigger than the sum of individual entropies. Equality (d) comes from definition of codewords described in Subsection \ref{codebook}. Equality (e) holds after subtracting off the dithers from outer codewords. Inequality (f) comes applying Crypto Lemma \cite{forney2004role} to the first term. Note that the jamming codewords are constructed from denser lattice sets and have uniform distribution over their codebooks, therefore the aligned codeword belongs to the same codebook and it is also uniformly distributed hence it is independent of the confidential message codewords. Inequality (g) is concluded from Lemma 1 in \cite{babaheidarian2015compute} and power allocation among message and jamming codewords described in Subsection \ref{encode}. Lastly, equality $h$ comes from the rate of constructed outer lattice codewords explained in Subsection \ref{codebook}.
\subsection{Proof of Corollary 1}\label{proof}
We present the proof in two steps. Step 1 is to prove that the first term in expression (\ref{theorem1eq}) achieves $\frac{1}{K}$ degrees of freedom for each user. Step 2 is to show that the second term in expression (\ref{theorem1eq}) is constant with respect to power constraint $P$ as power $P$ approaches infinity.\\
Step 1: Following Corollary 5 in \cite{ordentlich2014approximate}, all $K$ optimal combination rates in (\ref{theorem1eq}) offer $\frac{1}{K}$ degrees of freedom for almost every real-valued channel gain vector.\\
Step 2: This step is to show that for each user the second term in expression (\ref{theorem1eq}) is constant with respect to power $P$ as $P$ tends to large values, i.e., the following holds:
\begin{multline}
\lim_{P \rightarrow \infty} \bigg [\log\left(\sum_{m \in S \atop \frac{|S|}{M} \rightarrow 1}^M \frac{h_{\ell i}^2 P_{\ell,m}+h_{K-\ell+1,i}^2P_{K-\ell+1,m}^J}{h_{K-\ell+1,i}^2P_{K-\ell+1,m^*}^J}\right) \times \bigg .\\ \bigg. \frac{1}{\log(1+P)} \bigg] \rightarrow 0 \\
\end{multline}
Note that partial powers $P_{\ell,m}$ and $P_{\ell,m}^J$ are allocated power to encode the $m$-th component of the confidential message $\ell$ and the jamming signal at Transmitter $\ell$, respectively. These powers are tuned to satisfy the power constraint $P$. Therefore, at each Transmitter $\ell$ and for all $m \in \{1,2, \dots, M\}$ we have $P_{\ell,m}=\alpha_{\ell,m}P$ and $P_{\ell,m}^J=\beta_{\ell,m}P$ for some positive factors $0<\alpha_{\ell,m}<1$ and $0<\beta_{\ell,m}<1$. Therefore, we have
\begin{equation*}
\frac{h_{\ell i}^2 P_{\ell,m}+h_{K-\ell+1,i}^2P_{K-\ell+1,m}^J}{h_{K-\ell+1,i}^2P_{K-\ell+1,m^*}^J}
\end{equation*}
\begin{equation*}
=\frac{P\left(h_{\ell i}^2\alpha_{\ell,m}+h_{K-\ell+1,i}^2\beta_{K-\ell+1,m}\right)}{P\left(h_{K-\ell+1,i}^2\beta_{K-\ell+1,m^*}\right)}
\end{equation*}
\begin{equation}\label{cor1constant}
=\frac{h_{\ell i}^2\alpha_{\ell,m}+h_{K-\ell+1,i}^2\beta_{K-\ell+1,m}}{h_{K-\ell+1,i}^2\beta_{K-\ell+1,m^*}}
\end{equation}
where all the terms in expression (\ref{cor1constant}) are considered constant with respect to power $P$ and hence, the following is deduced
\begin{multline}
\lim_{P \rightarrow \infty} \bigg[ \log \left( \frac{h_{\ell i}^2 \alpha_{\ell,m}+h_{K-\ell+1,i}^2 \beta_{K-\ell+1,m}}{h_{K-\ell+1,i}^2\beta_{K-\ell+1,m^*}} \right) \times \bigg. \\ \shoveleft{ \bigg. \frac{1}{\log(1+P)} \rightarrow 0 \bigg]} \\
\end{multline}
Also, note that in \cite{xie2014secure} it was shown that sum secure degrees of freedom of $1$ is the upper bound for an arbitrary $K$-user interference channel with confidential messages. Therefore, our achievable sum secure degrees of freedom is optimal. This concludes the proof of Corollary \ref{cor1}.  
\section{Conclusion}\label{conclude}
We provided a new achievable security scheme to transmit confidential messages over an asymmetric interference channel with arbitrary number of users ($K >2 $) so long as interference is within weak and moderately weak interference regimes. Our achievable scheme utilizes the nested lattice codebooks, i.i.d. repetitive codes, cooperative jamming, superposition coding, and the compute-and-forward decoding strategy. \\
We showed that following our scheme, users achieve secure rates which scale linearly with $\log(\mathrm{SNR})$ and a sum secure rate that is within constant gap of sum capacity. Our cooperative scheme shared among transmitters achieves the sum secure degrees of freedom of $1$ without any online communication among transmitters or using external helpers.


\bibliographystyle{IEEEtran}
\bibliography{bib_1}

\end{document}